      [1999/12/01 v1.4c Il Nuovo Cimento]
\documentclass{cimento}

\usepackage{graphicx}  
\title{Gluon polarization and jet production at STAR}
\author{P.~Djawotho\from{ins:x} for the STAR Collaboration}
\instlist{\inst{ins:x} Cyclotron Institute, Texas A\&M University,
College Station, TX 77843-3366, USA}
\begin{document}

\maketitle

\begin{abstract}
At RHIC kinematics, polarized jet hadroproduction is dominated by $gg$ and $qg$
scattering, making the jet longitudinal double-spin asymmetry, $A_{LL}$, sensitive
to gluon polarization in the nucleon. I will present STAR results of $A_{LL}$ from
inclusive jet and dijet measurements for the RHIC 2006 run totaling 5.5 pb$^{-1}$ of
integrated luminosity with 55\% beam polarization at center-of-mass energy 200 GeV.
I will also present preliminary results from the analyses of data from the 2009 run,
which collected a much larger sample (25 pb$^{-1}$ with 57\% polarization) at 200 GeV.
The results are compared with theoretical calculations of $A_{LL}$ based on various models
of the polarized gluon distribution function in the nucleon. The STAR data place
significant constraints on allowed theoretical models.
\end{abstract}

\section{Introduction}

The study of the internal spin structure of the proton is an integral part of the
Relativistic Heavy Ion Collider (RHIC) spin physics program. The polarized proton
collider is especially well suited to measure the polarized gluon contribution to
the proton spin. The Solenoidal Tracker at RHIC (STAR) experiment, with its large
acceptance, retains an advantage in accessing $\Delta g(x)$ via jet production.

The longitudinal spin sum rule dictates how the proton spin is constructed from the
spin and orbital momenta of its partonic constituents:
\begin{equation}
\frac{1}{2} = \frac{1}{2}\Delta\Sigma+\Delta G+L_z,
\end{equation}
where the quark polarization, $\Delta\Sigma \approx 0.3$, has been measured in
deep-inelastic scattering experiments. However, $\Delta G$, the gluon polarization,
and $L_z$, the parton orbital angular momentum, are still poorly
constrained~\cite{ref:Ashman}. RHIC stands to bring significant advances in the
mapping of $\Delta g(x)$.

RHIC collected data at 200 GeV center-of-mass energy in polarized proton-proton
collisions with integrated luminosity of 5.4 pb$^{-1}$ in 2006 and 25 pb$^{-1}$ in 2009.
The beam polarization was measured with Coulomb-nuclear interference (CNI) proton-carbon
polarimeters~\cite{ref:Jinnouchi} calibrated with a polarized atomic hydrogen-gas
target~\cite{ref:Okada}. The average beam polarization was 55\% in 2006 and 57\%
in 2009.

The STAR detector subsystems~\cite{ref:RHIC} relevant to jet analysis are the Time
Projection Chamber (TPC) immersed in a 0.5 T longitudinal magnetic field and used to
reconstruct charged particle tracks with pseudorapidity $|\eta|<1.3$. The Barrel
Electromagnetic Calorimeter (BEMC) with towers at $|\eta|<1$ and the Endcap
Electromagnetic Calorimeter (EEMC) with towers at $1<\eta<2$ were used to measure
neutral particles and for triggering with jet patches of fixed size
$\Delta\eta\times\Delta\phi=1.0\times 1.0$.
The Beam-Beam Counters (BBC) with $3.3<|\eta|<5.0$ and Zero-Degree Calorimeters (ZDC)
located $\sim$18 m downstream of the interaction point were used for monitoring relative
luminosities. A timing window imposed on the BBCs was used as part of the minimum bias
trigger requirement in 2006. All of these detectors cover full azimuth
$(\Delta\phi=2\pi)$.

\section{Analysis and results}

Jets were reconstructed using a midpoint-cone algorithm~\cite{ref:Blazey} with seed
transverse energy 0.5 GeV, split-merge fraction 0.5 and cone radius 0.7. Tracks and
towers were required to have a minimum transverse momentum of 0.2 GeV/$c$. $A_{LL}$
is formally defined as the ratio of the difference over the sum of cross sections
with opposite helicity states. It can be measured at RHIC with~\cite{ref:Bunce}:
\begin{equation}
A_{LL} = \frac{1}{P_1 P_2}\frac{N^{++} - R N^{+-}}{N^{++} + R N^{+-}},
\end{equation}
where $P_1$ and $P_2$ are the beam polarizations, $R=\mathcal{L}^{++}/\mathcal{L}^{+-}$
is the ratio of luminosities, and $N^{++}$ and $N^{+-}$ are the jet yields for equal
and opposite helicity beams.

Several important improvements over the 2006 run were realized, both before and
after the taking of the 2009 data. Overlapping jet patches were added to the trigger
and lower $E_T$ thresholds were adopted for both the BEMC and EEMC. These upgrades
helped increase trigger efficiency and reduce trigger bias. They resulted in a 37\%
increase in jet acceptance over the 2006 run. Upgrades in the data acquisition system,
DAQ1000, allowed STAR to record events at several hundred Hz during the 2009 run,
with only 5\% dead time for the jet data, compared with 40 Hz with 40\% dead time
during the 2006 run. The enhanced DAQ capability also allowed STAR to remove the BBC
coincidence requirement, which helped trigger more efficiently at high jet $p_T$.
Improvements in jet reconstruction were also implemented. The electromagnetic
calorimeters are $\sim$1 hadronic interation length thick. Many charged hadrons
deposit a MIP (minimum ionizing particle), while others shower and deposit a sizeable
fraction of their energy when passing through. The strategy adopted in analyses
through 2006 was to subtract a MIP from an EMC tower with a charged track passing
through. In the 2009 run, the total momentum of the charged track is subtracted
from the struck EMC tower. This significantly reduces the response to fluctuations
from charged hadron showering and reduces the average difference between jet energies
at the detector and particle level. The net benefit comes in the form of an improved
overall jet energy resolution of 18\%, compared to 23\% in the 2006 analysis.

\begin{figure}
\centering
\includegraphics[width=0.7\textwidth]{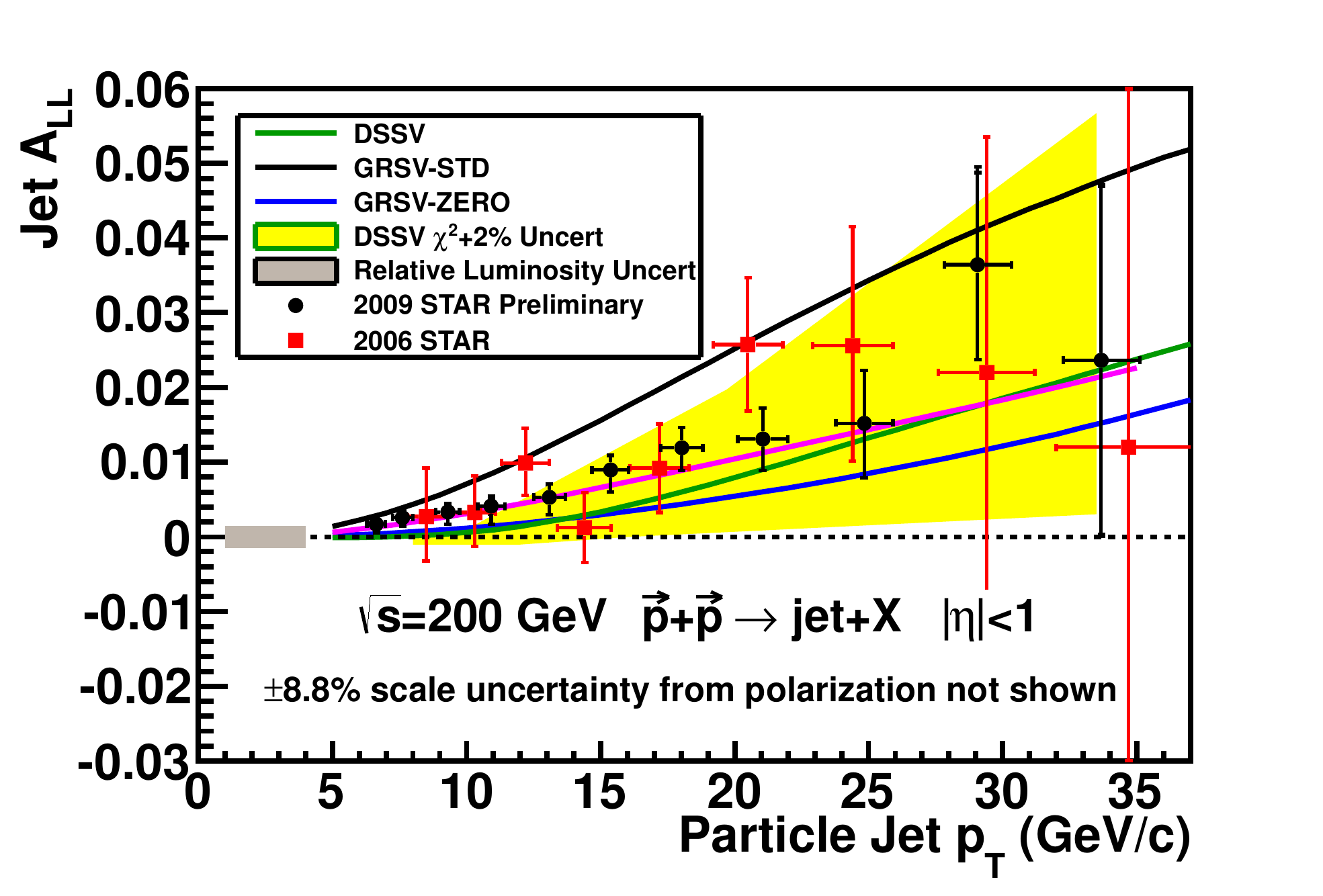}
\caption{STAR 2006 (red squares)~\cite{ref:Adamczyk} and 2009 (black circles) inclusive
jet $A_{LL}$ {\it vs.} jet $p_T$ for $|\eta|<1$~\cite{ref:Djawotho}.
The GRSV~\cite{ref:GRSV} model is a previous global analysis for polarized parton
densities that included DIS data. The GRSV-ZERO curve (blue) represents the special case
of no gluon polarization. The DSSV~\cite{ref:DSSV_PRL} model (green), in addition to DIS
and SIDIS data, is the first to incorporate RHIC polarized proton-proton data. The DSSV
$\chi^2+2\%$ curve~\cite{ref:DSSV_PPNP} (magenta) is described in the text.}
\label{fig:JetALL}
\end{figure}

\begin{figure}
\centering
\includegraphics[width=0.48\textwidth]{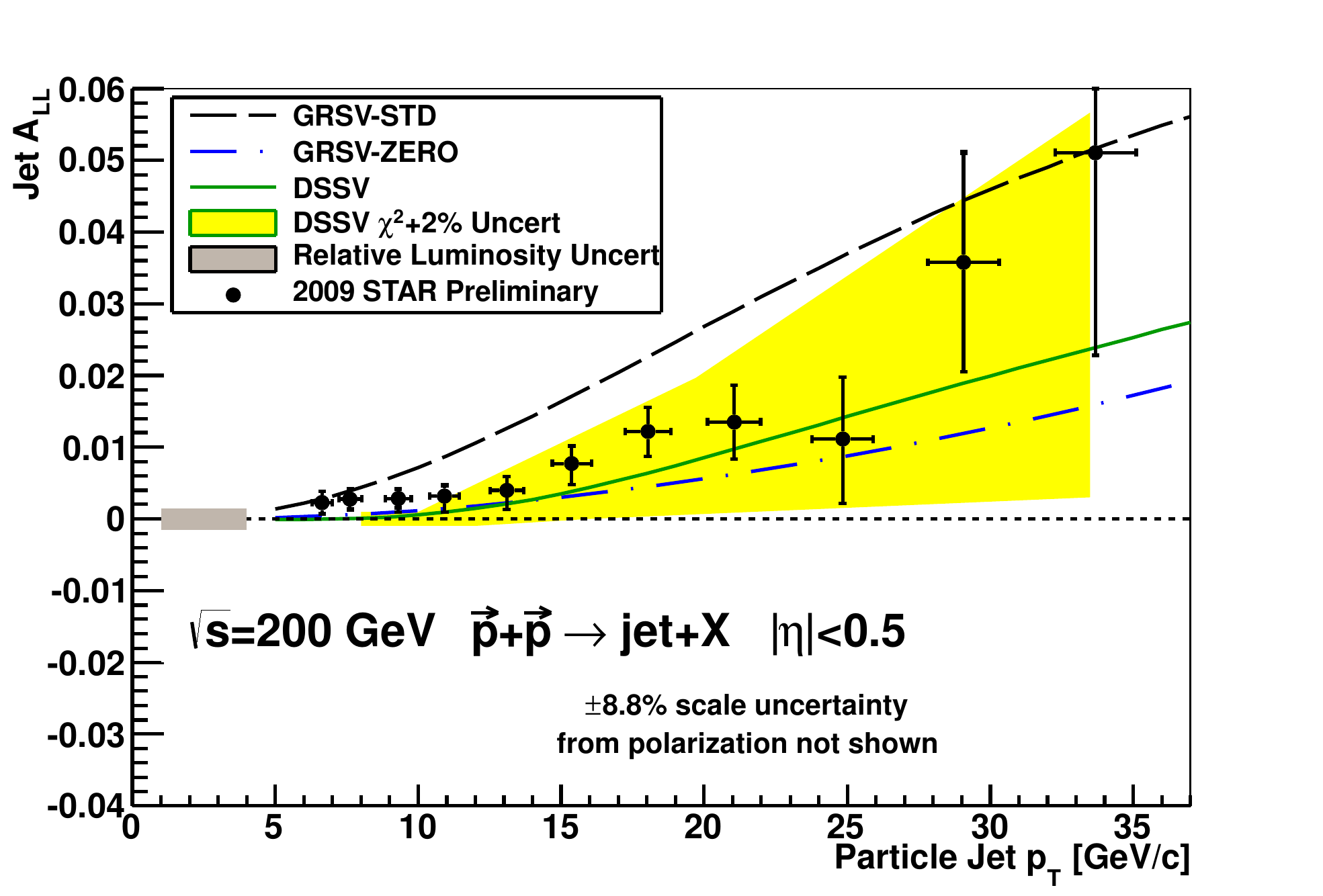}
\hfill
\includegraphics[width=0.48\textwidth]{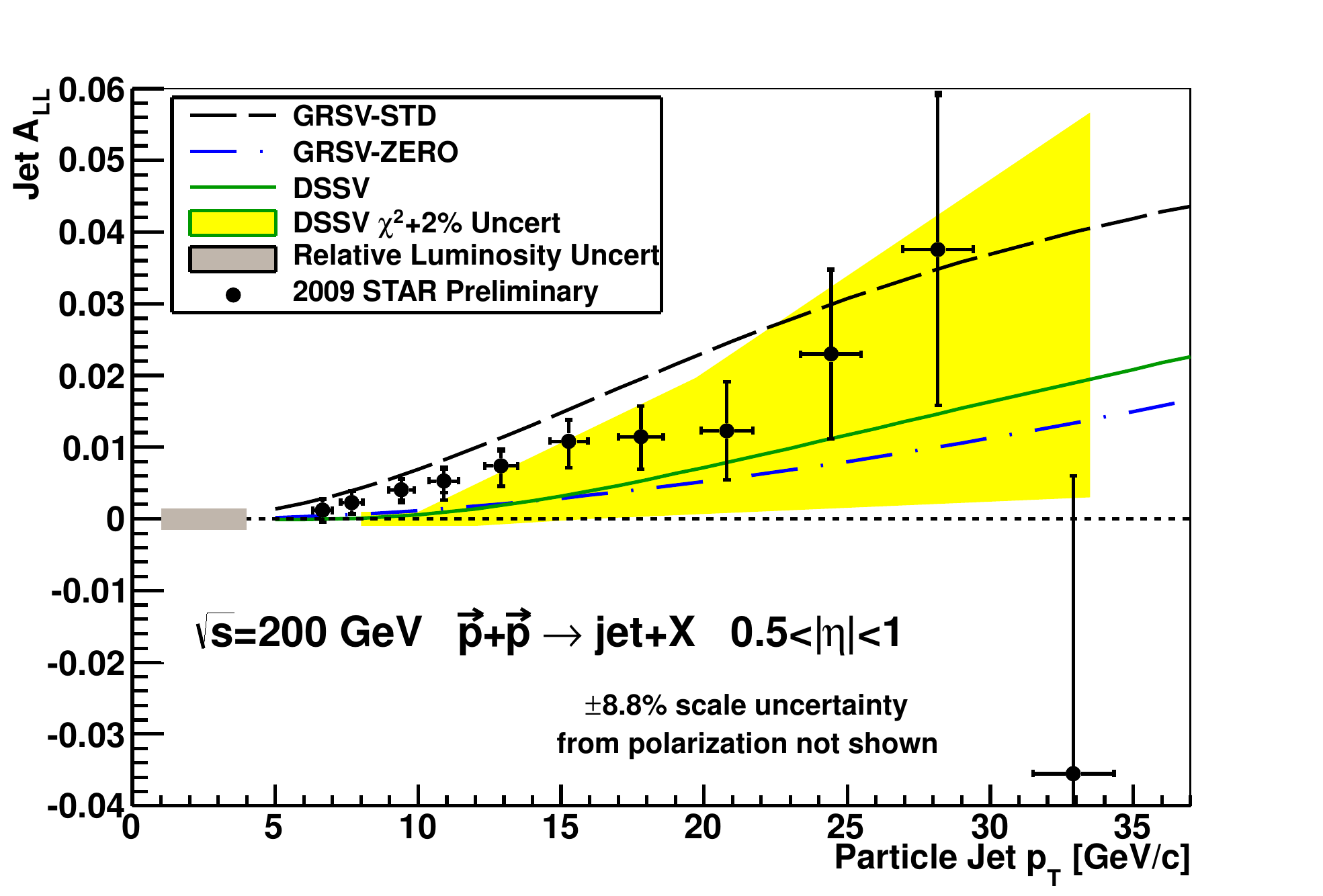}
\caption{STAR 2009 inclusive jet $A_{LL}$ {\it vs.} jet $p_T$ for the pseudorapidity ranges
$|\eta|<0.5$ (left panel) and $0.5<|\eta|<1$ (right panel).}
\label{fig:JetALLMidFwd}
\end{figure}

Figure~\ref{fig:JetALL} shows the measured inclusive jet $A_{LL}$ {\it vs.} jet $p_T$
for the 2006 $(-0.7<\eta<0.9)$~\cite{ref:Adamczyk} and 2009
$(|\eta|<1)$~\cite{ref:Djawotho} data alongside theory predictions of GRSV~\cite{ref:GRSV}
and DSSV~\cite{ref:DSSV_PRL}. The STAR data fall between the predictions of DSSV and
GRSV-STD. The dominant systematic uncertainties originate from differences between the
reconstructed and true jet $p_T$ and the trigger sampling the underlying partonic
processes ($qq$, $qg$ and $gg$) differently. The 2009 data are more precise than the 2006
data by a factor of four in low-$p_T$ bins and a factor of three in high-$p_T$ bins.
The magenta curve~\cite{ref:DSSV_PPNP} shows a fit to the previous datasets for which the
truncated integral of $\Delta g$ over the region $0.001<x<1$ was varied allowing the
$\chi^2$ of the fit to change by 2\%. It provides a very good description of the new 2009
results. The truncated integral of $\Delta g(x)$ over the range $0.05<x<0.2$ is
0.13~\cite{ref:DSSV_PPNP}. Figure~\ref{fig:JetALLMidFwd} presents the 2009 result in two
rapidity ranges, permitting comparisons with models for collisions with different average
partonic scattering angles, $x$ ranges and subprocess mixtures.

\begin{figure}
\centering
\includegraphics[width=0.9\textwidth]{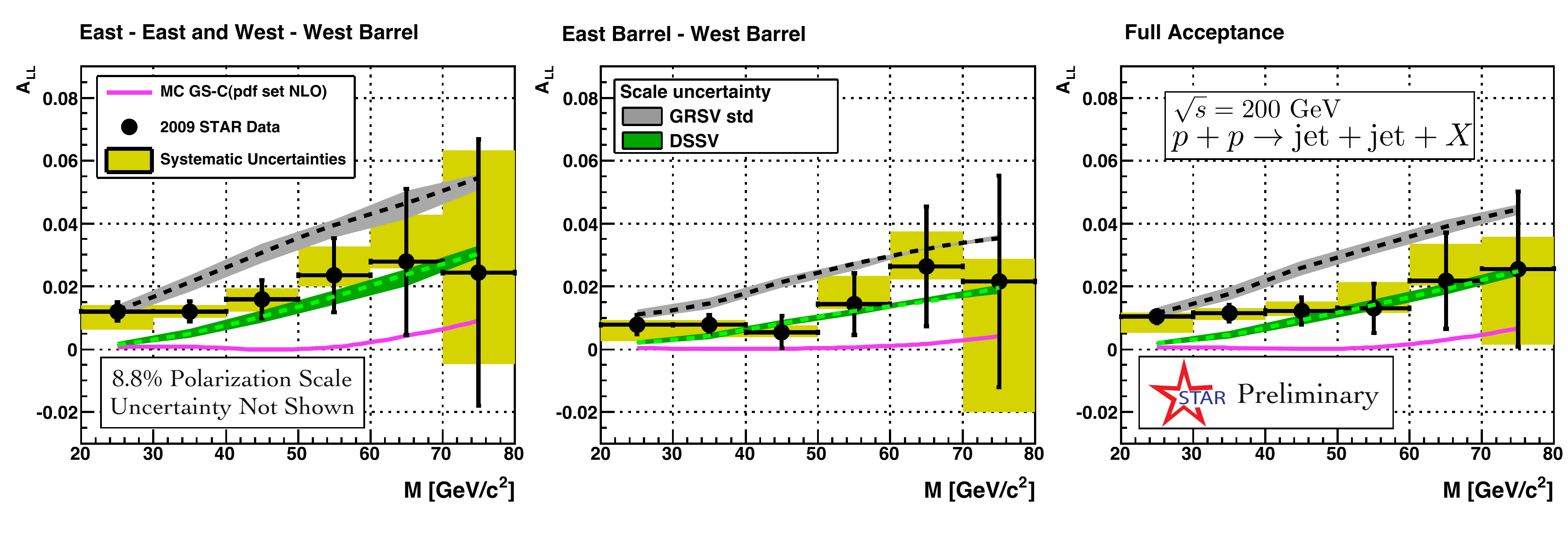}
\caption{STAR 2009 dijet $A_{LL}$~\cite{ref:Walker} measured in three pseudorapidity
acceptances to better constrain the kinematics of the hard-scattering partons.}
\label{fig:DijetALL}
\end{figure}

Figure~\ref{fig:DijetALL} shows the measured dijet $A_{LL}$~\cite{ref:Walker} compared
with the predictions of GRSV~\cite{ref:GRSV}, DSSV~\cite{ref:DSSV_PRL} and
GS-C~\cite{ref:GSC}. The dijet results, which also fall between the DSSV and GRSV
predictions, will help constrain the shape of $\Delta g(x)$.

\section{Conclusion}

The STAR experiment measured the inclusive jet double-helicity asymmetry $A_{LL}$ in
polarized proton-proton collisions at $\sqrt{s}=200$ GeV.
The STAR 2005 and 2006 measured inclusive jet $A_{LL}$ were included in the first
global analysis~\cite{ref:DSSV_PRL} to use polarized jets and played a significant
role in constraining $\Delta g(x)$ at RHIC kinematics. The markedly increased
precision of the 2009 result is expected to vastly reduce the present large
uncertainty of the gluon polarization of the proton once included in a global
analysis of polarized parton densities. The complementary STAR measurement of $A_{LL}$
for dijets from 2009 will reduce the uncertainty in $\Delta g(x)$ associated with
extrapolating beyond the $x$ range explored by the inclusive jet measurement.


\begin{thebibliography}{99}
\bibitem{ref:Ashman} \BY{Ashman~J. {\it et al.}}
                     \IN{Nucl. Phys. B}{328}{1989}{1};
                     \BY{Filipone~B.W. \atque Ji~X.D.}
                     \IN{Nucl. Phys.}{26}{2001}{1}.
\bibitem{ref:Jinnouchi} \BY{Jinnouchi~O. {\it et al.}}
                        arXiv:nucl-ex/0412053.
\bibitem{ref:Okada} \BY{Okada~H. {\it et al.}}
                    arXiv:hep-ex/0601001.
\bibitem{ref:RHIC} \TITLE{Special Issue on RHIC and its Detectors},
                   edited by \BY{Harrison~M., Ludlam~T. \atque Ozaki~S.}
                   \IN{Nucl. Instrum. Methods Phys. Res., Sect. A}{499}{2006}{624}.
\bibitem{ref:Blazey} \BY{Blazey~G.C. {\it et al.}}
                     arXiv:hep-ex/0005012.
\bibitem{ref:Bunce} \BY{Bunce~G., Saito~N., Soffer~J. and Vogelsang~W.}
                    \IN{Annu. Rev. Nucl. Part. Sci.}{50}{2000}{525}.
\bibitem{ref:Adamczyk} \BY{Adamczyk~L.}
                       \IN{Phys. Rev. D}{86}{2012}{032006}.
\bibitem{ref:Djawotho} \BY{Djawotho~P.}
                       arXiv:1106.5769.
\bibitem{ref:GRSV} \BY{Gl\"{u}ck~M., Reya~E., Stratmann~M. \atque Vogelsang~W.}
                   \IN{Phys. Rev. D}{70}{2004}{034010}.
\bibitem{ref:DSSV_PRL} \BY{de~Florian~D., Sassot~R., Stratmann~M. \atque Vogelsang~W.}
                       \IN{Phys. Rev. Lett.}{101}{2008}{072001}.
\bibitem{ref:DSSV_PPNP} \BY{de~Florian~D., Sassot~R., Stratmann~M. \atque Vogelsang~W.}
                        \IN{Prog. Part. Nucl. Phys.}{67}{2012}{251}.
\bibitem{ref:Walker} \BY{Walker~M.}
                     arXiv:1107.0917.
\bibitem{ref:GSC} \BY{Gehrmann~T. \atque Stirling~W.J.}
                  \IN{Phys. Rev. D}{53}{1996}{6100}.
\end{thebibliography}
\end{document}